\documentclass{nature}

\usepackage{amsmath,bm,color,placeins}
\usepackage[version=3]{mhchem}
\usepackage{graphicx}
\makeatletter
\let\saved@includegraphics\includegraphics
\AtBeginDocument{\let\includegraphics\saved@includegraphics}
\renewenvironment*{figure}{\@float{figure}}{\end@float}
\makeatother

\title{Probing Phase Diagrams of Ordered Two-Dimensional Ice}

\author{Bingzheng Wu$^{1}$, Jianming Wu$^{1}$, Sai Duan$^{1,2\ast}$,
\& Xin Xu$^{1,2\ast}$}

\begin{document}

\maketitle

\begin{affiliations}
\item State Key Laboratory of Porous Materials for Separation and
Conversion, Shanghai Key Laboratory of Molecular Catalysis and
Innovative Materials, MOE Key Laboratory of Computational Physical
Sciences, Research Center for Chemical Theory, Department of Chemistry,
Fudan University, Shanghai 200438, P.~R. China
\item Hefei National Laboratory, Hefei 230088, P.~R. China
\end{affiliations}

\begin{abstract}
Water, a ubiquitous and fundamental substance, plays a critical role
across a wide range of disciplines from physics and chemistry to biology
and engineering. Despite theoretical predictions of several phases of
two-dimensional (2D) ice confined between idealized hydrophobic walls,
experimental validation has been limited to the square phase, whose
structural origin remains controversial. Here, we propose a realistic
nanoconfinement setup using wide-bandgap hexagonal boron nitride (h-BN)
as the capping layer and Cu(111) as the substrate. This protocol enables
scanning tunneling microscope (STM) to resolve the atomic-scale
arrangement of water molecules beneath the h-BN layer, overcoming the
limitations of conventional techniques. Simulated STM images
unambiguously identify all ordered flat 2D ice phases, as well as
coexisting phases, and effectively distinguish them from potential
contaminants. These findings establish a robust framework for experiment
to systematically probe the phase structures of 2D ice, opening an
avenue for studying nanoconfined water under ambient conditions.
\end{abstract}

\newpage


Water is the most essential chemical substance for life on our
planet\cite{SzentGyrgyi1971,ball2017pnas}. One of the remarkable
characteristics of bulk water is its array of anomalous physical
properties\cite{ball2008nature,pettersson2016cr}, which are
intrinsically linked to its diverse phase 
structures\cite{eisenberg2005,brini2017cr}. Consequently, exploration of
the water's phase diagrams has been a central theme for over a
century\cite{tammann1900ap,Hansen2021,salzmann2023science}. When water
is confined to lower dimensions, conditions prevalent in interfacial
systems, its behavior becomes even more intriguing. For instance, water
molecules confined in nanoslits exhibit a range of unusual
physicochemical properties, such as peculiar ion
permeability\cite{Gopinadhan2019}, anomalous low dielectric
constant\cite{Fumagalli2018} and friction coefficient\cite{Nair2012}, as
well as significantly enhanced shear viscosity\cite{Neek-Amal2016} and
thermal conductivity\cite{Guo2025}. As in the bulk, efforts have been
made to correlate these anomalous behaviors with structural features
through proposed 
hypotheses\cite{Israelachvili1996,Gopinadhan2019,Fumagalli2018} and
theoretical models\cite{Kavokine2022}. However, in low-dimensional
systems, definitive structure-property relationships, comparable to
those established for three-dimensional water, remain elusive.

This challenge primarily stems from the lack of chemically resolved
atomic-scale characterization techniques to determine the phase
structures of confined water. Although several structures without
confinement, such as buckled pentagonal rings coexisting with
heptagonal\cite{Nie2010} and hexagonal arrangements\cite{Ma2020}, have
been observed, fully studying the phase diagrams of two-dimensional (2D)
ice requires confinement to generate the necessary pressure, which
inevitably imposes limitations on the applicability of conventional
measurement techniques\cite{Yang2009,Kimmel2009,Xu2010,He2012,Li2015}.
As a result, compared to bulk, experimental approaches on structural
characterization of nanoconfined 2D ice remain scarce.

\begin{figure}[!tbh]
\centering
\includegraphics[width=\textwidth]{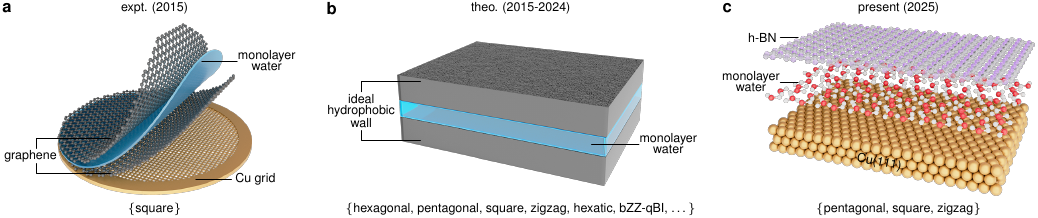}
\caption{\textbf{Configurations for 2D ice phases}. \textbf{a},
A water monolayer confined between two graphene monolayers, with the
bottom sheet supported on a Cu grid. With this setup, experimental
measurements via transmission electron microscopy (TEM) speculated a
square arrangement of the water molecules. \textbf{b}, Hydrophobic walls
were used in theoretical simulations to explore phase diagrams of
nanoconfined 2D ice. This idealized configuration approximates the van
der Waals interactions between nanoslits and water molecules but does
not correspond to a realistic experimental setup. \textbf{c}, The setup
proposed in this work, where a water monolayer is confined between a
realistic Cu(111) substrate and a scanning tunneling microscope 
(STM)-transparent h-BN capping layer. The publication years of the
respective studies are labeled in parentheses, and the proposed or
observed phases are labeled in the braces.}
\label{fig1}
\end{figure}

Algara-Siller et al. pioneered the experimental investigation of 2D ice
under lateral pressure by confining water between two graphene sheets
(Figure~\ref{fig1}a)\cite{AlgaraSiller2015}. Using transmission electron
microscopy (TEM)\cite{AlgaraSiller2015}, they observed an ordered
``square'' ice structure at a lateral pressure of approximately 1~GPa,
opening the door for the study of nanoconfined ice. Subsequent
theoretical studies, which employed idealized hydrophobic walls as the
confinement layers (Figure~\ref{fig1}b), revealed a rich phase diagram
of 2D water. Specifically, Chen et al. successfully reproduced the
experimentally observed ``square'' structure and identified additional
arrangements, such as pentagonal and buckled rhombic arrangements under
varying lateral pressures and confinement widths at 0~K\cite{Chen2016}.
By further considering temperature effects, theoretical phase diagrams
of 2D ice were subsequently 
proposed\cite{Kapil2022,Lin2023,jiang2024nphys,Ravindra2024}, indicating
that these ordered structures remain stable over a wide range of
temperatures from 0 to 170~K. 

Despite a variety of 2D ice structures predicted by theory,
direct comparison between theoretical models and experimental
observations remains challenging due to the exclusive use of idealized
hydrophobic confinement potentials and a primary reliance on energetic
calculations\cite{Chen2016,Kapil2022,Lin2023,jiang2024nphys,Ravindra2024}.
In fact, to date, the only experimentally observed arrangement remains
the initially reported square structure\cite{AlgaraSiller2015}. However,
the precise identification of this structure is still under debate
because of the lack of chemical resolution of TEM imaging. Zhou et al.
argued that sodium chloride (NaCl), a common contaminant in graphene,
cannot be excluded, as it exhibits a similar structure and lattice
constant that cannot be distinguished from the 2D square ice by
TEM\cite{Zhou2015}. Therefore, to eliminate plausible candidates with
similar geometrical parameters and thus to establish definitive evidence
for the structures of nanoconfined 2D ice, a characterization technique
that combines both spatial and chemical resolution is urgently needed.

In this work, we propose a setup that enables the leverage of the
chemical sensitivity of scanning tunneling microscope (STM) to
characterize different phase structures of 2D ice (Figure~\ref{fig1}c).
In this protocol, the Cu(111) surface is selected as the substrate
because of its inherent hydrophobicity and high electrical
conductivity\cite{duan2020jacs}. Meanwhile, hexagonal boron nitride
(h-BN) is chosen as the capping layer, which provides the required
lateral pressure for stabilizing various 2D ice phases while remaining
STM-transparent owing to its 2D structure and wide
bandgap\cite{Wickramaratne2018,Wu2019,Kirchhoff2022}. By adjusting the
confinement width between h-BN and Cu(111), lateral pressures applied to
2D ice can be effectively controlled\cite{Chen2016}. Simulations based
on this setup\cite{Tersoff1985,duan2020jacs,duan2023jacsau,duan2023jpcl,zhu2025jacsau}
demonstrate that STM measurements can clearly distinguish
not only between different ordered ice phases but also between these
phases and potential contaminants such as NaCl. Hence, this approach
provides a unique and robust platform that all ordered phase structures
of 2D ice can be unambiguously identified and explored, for the first
time.

\section*{\large{Results and discussion}}
\subsection*{\textbf{Identifying the square phase of 2D ice}} 
\begin{figure}[!tbh]
\centering
\includegraphics{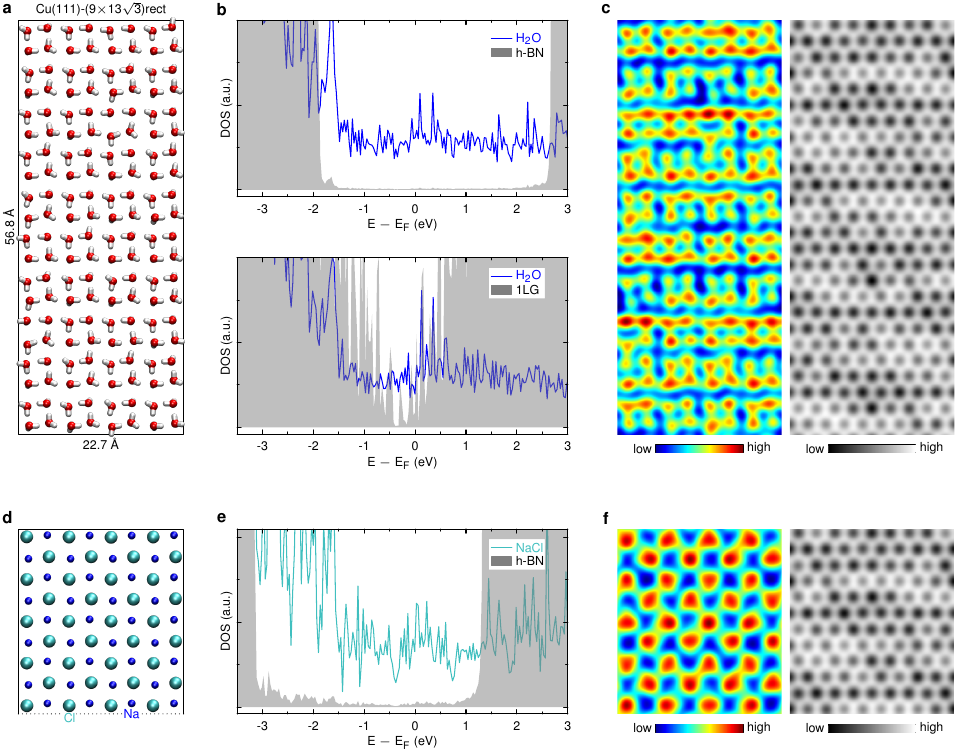}
\caption{\textbf{Identification of the square 2D ice}. \textbf{a}, Top
view of the optimized geometry of the square 2D ice confined between a
h-BN layer and a Cu(111) surface, with a confinement width of 5.7~\AA{}.
The lattice constants (in~\AA{}) of the heterogeneous layers and the
supercell of Cu(111) are labeled. \textbf{b}, Projected density of state
(DOS) for the square 2D ice system with a single h-BN layer (top) and a
single graphene layer (1LG, bottom) as the capping layer. \textbf{c},
Simulated STM images for confined square 2D ice with h-BN (left) and
graphene (right) as the capping layer. \textbf{d}, Top view after
replacing the water layer in panel (\textbf{a}) with a NaCl layer.
\textbf{e}, Projected DOS for the system of NaCl capped by h-BN.
\textbf{f}, Simulated STM images of confined NaCl with a single h-BN
layer (left) and a single graphene (right) as the capping layer. All STM
simulations were performed at a sample bias of --0.5~V.}
\label{fig2}
\end{figure}

\\
We begin by examining the only experimentally observed ordered
arrangement, namely the ``square'' structure\cite{AlgaraSiller2015}. The
optimized geometry of this square 2D ice, confined between h-BN and
Cu(111) at a confinement of 5.7~\AA{}, is depicted in Figure~\ref{fig2}a.
The primary structural feature of this 2D ice, i.e., square units formed
by four oxygen atoms at the vertices, agrees well with previous
experimental\cite{AlgaraSiller2015} and theoretical
results\cite{Chen2016,Kapil2022}, despite that the latter adopted
idealized hydrophobic repulsive wall potentials. Notably, due to the
presence of the realistic Cu(111) substrate and h-BN capping layer, a
slight structural distortion is observed with some \ce{O-H} bonds
pointing towards the metallic substrate. The calculated lateral pressure
is 0.9~GPa, well within the previously reported range of 0.6--1.7~GPa
from simulations\cite{Kapil2022}. Therefore, the minor structural
deviations induced by inevitably water-substrate interactions in
realistic conditions do not preclude the identification of this
configuration as the square 2D ice phase.

\FloatBarrier

The calculated density of states (DOS) reveals that the contribution
from the top h-BN layer near the Fermi level is negligible
(Figure~\ref{fig2}b, top panel), attributable to its wide bandgap of
around 4~eV\cite{Wickramaratne2018,Wu2019,Kirchhoff2022}. In contrast,
the DOS contribution from the water layer is two orders of magnitude
higher than that of h-BN, indicating that the underlying 2D ice
structures can be readily detected by STM. Indeed, the simulated STM
image shows distinct protrusions corresponding to oxygen atoms beneath
the h-BN layer (Figure~\ref{fig2}c, left panel), confirming that the STM
transparency of h-BN enables chemically resolved imaging of the 2D ice.
It is noted that the STM pattern is slightly non-uniform with fused
oxygen atoms in some rows. This heterogeneity reflects the influence of
lattice mismatch between the 2D ice and the metallic substrate.

To highlight the STM transparency of h-BN, we further simulated the same
confined ``square'' ice structure between a graphene layer and Cu(111),
a configuration that closely resembles the experimental setup used in
TEM study\cite{AlgaraSiller2015}. Due to the unique band structure of
the Dirac cone\cite{Geim2007,Abergel2010}, the DOS near the Fermi level
is dominated by the graphene layer, despite a slight increase in the DOS
contribution from the water molecules (Figure~\ref{fig2}b, bottom panel).
As a result, STM measurements cannot effectively probe through the
graphene capping layer and are thus unable to resolve the structural
arrangement of the underlying 2D ice. Instead, only the characteristic
honeycomb pattern of graphene is observed in the simulated STM image
(Figure~\ref{fig2}c, right panel). This contrast underscores the
critical importance of using wide bandgap h-BN as the capping layer to
enable STM transparent confinement, thereby allowing chemically resolved
imaging of the structures beneath the capping layer, which is
inaccessible in other scanning probe 
techniques\cite{gross2009science,duan2015jacs,duan2016angew,duan2019jacs,qiu2022jacsau}.

Another key advantage of the current setup is its capability to identify
the debated contaminants\cite{AlgaraSiller2015,Zhou2015}. To this end,
we replaced the water molecules in the square 2D ice with sodium and
chlorine atoms in a 1:1 stoichiometric ratio, simulating a potential
NaCl contaminant\cite{Zhou2015} that may be introduced during the
transfer of 2D materials or from environmental pollutions
(Figure~\ref{fig2}d). In this scenario, the bandgap of h-BN remains
intact but exhibits a significant negative band offset
(Figure~\ref{fig2}e). Accordingly, h-BN again exhibits negligible DOS
near the Fermi level. Meanwhile, the DOS of the NaCl layer is comparable
to that of the square 2D ice. Consequently, the structure of the
confined NaCl is also detectable via STM. Interestingly, due to the
dominant contribution of the Cl atoms (Extended Data
Figure~\ref{fig_ex1}), the corresponding STM image has a significantly
(around 1.4~times) larger lattice of the square unit cell with a
45$^\circ$ rotation compared to that of the 2D ice (Figure~\ref{fig2}f,
left panel). These distinct structural and electronic signatures enable
unambiguous differentiation between square 2D ice and NaCl, helping to
resolve a long-term controversy in this subject. Again, when graphene is
used as the capping layer, the Dirac cone (Supplementary Figure~1)
causes the honeycomb pattern to dominate the corresponding STM image,
masking any signal from the beneath NaCl layer (Figure~\ref{fig2}f,
right panel).

Notably, the square arrangement of 2D ice is robust. For instance, when
the primitive water unit cell reported in Ref.~\citenum{Lin2023} is used,
the aspect ratio of the supercell is significantly decreased.
Nevertheless, the confinement width is only slightly reduced from 5.7 to
5.5~\AA{} to maintain the required lateral pressure of 1.0~GPa for
stabilizing the square arrangement. Simulated STM images reveal that, in
this case, the underneath water layer and the NaCl contaminant can again
be unambiguously distinguished when an h-BN capping layer is used,
whereas these features become indistinguishable when a graphene capping
layer is employed (Extended Data Figure~\ref{fig_ex2}). This result
further underscores the advantage of using the STM-transparent h-BN
capping layer, as discussed above.

\subsection*{\textbf{Probing other phases of 2D ice}}
\begin{figure}[!tbh]
\centering
\includegraphics{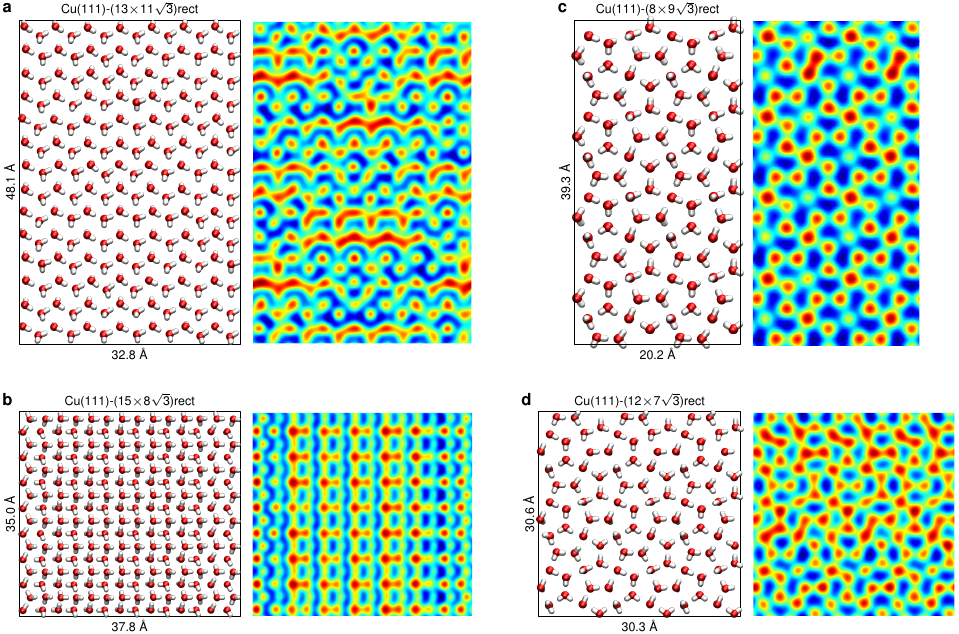}
\centering
\caption{\textbf{Probing other 2D ice structures}. Top view of the
optimized geometries (left) and the corresponding simulated STM images
(right) for the zigzag (\textbf{a}), square/zigzag coexisting
(\textbf{b}), pentagonal with rectangle supercell (\textbf{c}), and
pentagonal with square supercell (\textbf{d}) 2D ice confined between
the h-BN and Cu(111) surface. The lattice constants (in~\AA{}) of the
heterogeneous layers and the supercell of Cu(111) are labeled. All STM
simulations were performed at a bias voltage of --0.5 V.}
\label{fig3}
\end{figure}

\\
The present setup is also effective for identifying other ordered 2D ice
phases. Specifically, when the confinement width between h-BN and Cu(111)
is set to 4.8~\AA{}, a nanoconfined ``flat rhombic'' 2D
ice\cite{Kapil2022} is obtained, where the water arrangement displays
zigzag rows and linear columns, under the predicted lateral pressure of
2.6~GPa (Figure~\ref{fig3}a, left panel). The corresponding simulated
STM image displays pronounced stripe-like features along the rows, with
point-like features along the columns (Figure~\ref{fig3}a, right panel).
While the column features resemble those of the square phase, the row
features are quite different. This discrepancy can be attributed to the
strengthened hydrogen bonding within the zigzag water chains under
higher lateral pressure\cite{Ravindra2024}.

This STM imaging protocol not only captures the stable 2D ice structures,
but also enables clear visualization of the key intermediate structures
formed during phase transitions, thus providing insights into the phase
transformation mechanisms of 2D ice. Previous molecular dynamics
simulations employing idealized hydrophobic potentials suggested a broad
coexistence region between zigzag (flat rhombic) and square
phases\cite{Kapil2022,Lin2023}. Under the h-BN/Cu(111) nanoconfinement,
such a coexistence structure is observed at a confinement width of
6.5~\AA{} and a lateral pressure of 3.8~GPa (Figure~\ref{fig3}b, left
panel), where the square and zigzag arrangements located in the central
and edge regions, respectively. The corresponding STM patterns
faithfully reflect the coexistence of ordered water phases. Specifically,
rectangular units in the center gradually evolve into rhombic and
stripe-like patterns toward the edges (Figure~\ref{fig3}b, right panel).
Compared to the individual square and zigzag phases, the STM image of
the coexisting phase shows alternating rows with periodic bright-dark
contrast modulations. This contrast arises from differences in
water-substrate interactions across the interface. Specifically, the
excessively high surface density induces strong repulsive interactions
between adjacent water rows, leading to buckling and dislocation along
the surface normal direction in the 2D ice (Supplementary Figure~2). As
a result, the water molecules closer to the metallic substrate dominate
the STM contrast, as their enhanced interaction shifts their DOS closer
to the Fermi level. This highlights the power of STM imaging in
resolving the the subtle structural and electronic variations during
phase transitions in confined 2D ice.

Under lower lateral pressures, which are typically associated with
larger confinement widths, the STM-based detection generally faces
greater challenges in identifying the underneath water structures, due
to exponentially decaying tunneling currents and reduced structural
order. To test the effectiveness of the protocol proposed here, under
such demanding conditions, we simulated nanoconfined pentagonal 2D ice
at low lateral pressures ranging from 0.1 to 0.2~GPa. Previous
theoretical simulations indicated that there are two pentagonal phases,
differing in the shape of their unit cells. For the one with a
square-like unit cell reported in Ref.~\citenum{Kapil2022}, the optimal
confinement width between h-BN and Cu(111) is determined as 5.6~\AA{}.
The corresponding STM image displays a clearly visible pentagonal
arrangement of the beneath water molecules, as anticipated
(Figure~\ref{fig3}c). Notably, this confinement width is slightly
smaller than that (5.7~\AA{}) required for the square phase, which is
denser. This reflects the influence of the realistic confinement
conditions in the practical simulations. When the one with a
rectangle-like shaped unit cell from Ref.~\citenum{Chen2016} was adopted,
the optimized confinement width increases to the largest value of
6.8~\AA{}. Even under such weak confinement, STM measurements can still
effectively penetrate the h-BN layer, enabling unambiguous visualization
of the underlying pentagonal structure (Figure~\ref{fig3}d). In this
case, the contrast variation in the STM image arises again from subtle
height differences in the 2D ice lattice, induced by the realistic
water-metal interaction (Supplementary Figure~3).

It is worth reiterating that the key to the success of the current setup
in detecting the underneath water arrangements lies in our choice of
using h-BN as the capping layer, which has a negligible DOS distribution
near the Fermi level (Figure~2 and Extended Data Figure~\ref{fig_ex3}).
For systems with larger confinement widths, the DOS of h-BN decreases
further, which enhances the STM transparency of the capping layer and
effectively compensates for the signal attenuation typically associated
with increased tip-sample distance. Consequently, chemically resolved
images of the underlying water molecules can be obtained for all
investigated phases of 2D ice. In contrast, although graphene is widely
used to construct confined environments for stabilizing various 2D ice
phases, its unique Dirac cone band structure (Supplementary Figure~4)
results in a finite and delocalized DOS near the Fermi level. This
significantly masks the local electronic signatures of the encapsulated
water structures, making it difficult to detect the underneath 2D water
structures (Supplementary Figure~5).

\begin{figure}[!tbh]
\centering
\includegraphics{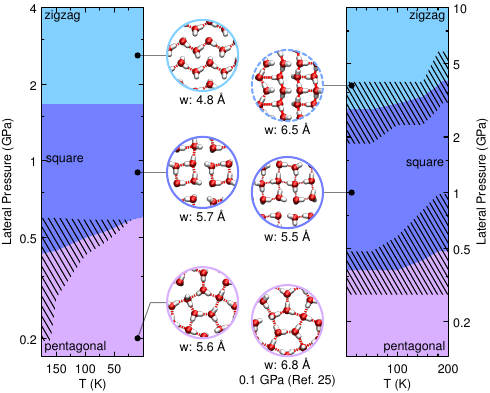}
\caption{\textbf{Correlation between the confinement widths and the
lateral pressures}. Summary of the calculated lateral pressures with
different confinement widths (labeled values in~\AA{}) for the 2D ice
confined between a single h-BN capping layer and Cu(111) substrate.
These results are generally in good agreement with the previously
predicted phase diagrams from Ref.~\citenum{Kapil2022} (left) and
Ref.~\citenum{Lin2023} (right) with idealized hydrophobic repulsive
potentials. The pentagonal phase with a confinement width of 6.8~\AA{}
corresponds to the lattice constants in Ref.~\citenum{Chen2016}. The
dashed areas in the phase diagrams indicate regions of coexistence
phases.}
\label{fig4}
\end{figure}

For the substrate, a real Cu(111) is used in the current setup, which
differs from the idealized confining environments used in previous
theoretical studies. Nevertheless, the obtained trend of lateral
pressure for various 2D ice phases is in good agreement with prior
reports (Figure~\ref{fig4}). For instance, using the lattice parameters
from Refs.~\citenum{Kapil2022} and \citenum{Lin2023}, we find that
reducing the interlayer spacing between h-BN and Cu(111) generally leads
to higher lateral pressure, and thus denser ice phases (Figure~\ref{fig4}
and Supplementary Table~1). The calculated lateral pressures exhibit
only minor variations when other 2D material capping layers, such as
graphene (Supplementary Table~2), are used, which highlights the
generality of the simulated pressure-confinement relationship. The most
notable exception is the coexistence phase between zigzag and square 2D
ice structures. While the predicted lateral pressure (3.6~GPa) for this
phase matches the value reported in Ref.~\citenum{Lin2023}, the
corresponding confinement width is unexpectedly larger than those of
lower-density phases. This anomalous behavior suggests that additional
external pressurization may be necessary to stabilize this phase under
real experimental conditions. It should be stressed that the
configurations identified here encompass all known ordered flat 2D ice
phases, which also represents the 2D ice structures over a wide range of
temperatures below 170~K (Figure~\ref{fig4}). However, the detection of
other proposed 2D ice phases, such as the zigzag quasi-bilayer or
hexatic ice\cite{Kapil2022,Lin2023}, under extreme conditions (e.g., high
temperature or ultrahigh pressure) remains an open challenge. Their
observation is hindered by the strong interaction between water and
confinement materials, as well as the intrinsic structural disorder,
necessitating the development of advanced detection techniques beyond
current capabilities.

Finally, we emphasize that the protocol proposed in the present work is
fully compatible with current experimental techniques. For example,
nanoslits can be pre-fabricated using electron beam lithography,
followed by transfer of h-BN flake onto the
structure\cite{pagliero2025nl}. Water molecules can then be introduced
through designed opening windows either at the ends\cite{pagliero2025nl}
or from the backsides\cite{Fumagalli2018}. Subnanometer-confined water
molecules can be achieved\cite{Fumagalli2018}, provided that sagging of
the h-BN capping layer during water infiltration can be
prevented\cite{pagliero2025nl}. Alternatively, a recently developed
approach involves first adsorbing water molecules onto a substrate under
controlled humidity and temperature, followed by direct transfer of an
h-BN flake to encapsulate the layer\cite{zheng2024arxiv}. This method
has successfully produced subnanometer-confined water molecules between
diamond and h-BN, offering precise control over confinement
thickness\cite{zheng2024arxiv}. Notably, a closely analogous nanoslit, featuring
a CO monolayer intercalated between Pt(111) and h-BN, has already been
realized using a protocol similar to the first approach\cite{Wu2019}.
This experimental demonstration strongly supports the feasibility of the
present design: a molecular layer confined between a metallic surface
and a h-BN capping layer, as proposed in the current work.

\section*{\large{Conclusion}}
In summary, this work proposes a unique protocol, using STM to
investigate the phases of the ordered 2D ice. In contrast to the
idealized hydrophobic potentials used in previous studies, it consists
of a wide bandgap h-BN as the capping layer and the Cu(111) surface as
the substrate. First-principles theoretical calculations reveal that STM
measurements can penetrate the h-BN layer, enabling chemically resolved
characterization of the underneath water structures, a capability
inaccessible to other experimental techniques. Consequently, the STM
images faithfully reflect the molecular arrangement of ``square'' ice
previously inferred from TEM studies, while clearly distinguishing it
from the potential NaCl contaminant, resolving a long-standing debate in
the field of 2D ice. Moreover, this protocol allows unambiguous
identification of all other previously predicted ordered 2D ice phases,
including the coexisting phase under various lateral pressures. This
makes STM a practical and powerful tool for probing 2D ice phases
over a wide range of lateral pressures. These findings not only deepen
our understanding of the microscopic structure of monolayer water but
also offer valuable insights into interactions between water and metal
surfaces in confined environments, lifting the curtain on the study of
phases of 2D ice at the atomic scale..

\section*{\large{Methods}}
\textbf{Model construction}\\
The slab model was employed to simulate the 2D heterogeneous structures,
where the metallic substrate was mimicked by a three-layer Cu(111) slab.
To avoid artificial deformation, the 2D ice and Cu(111) unit cells were
expanded such that the lattice mismatch ratio between the two supercells
is less than 1\%. A vacuum region exceeding 15~\AA{} was added above the
capping layer to render the inter-slab interactions along the
$z$-direction negligible.

\noindent\textbf{Structure optimization and lateral pressure evaluation}\\
The initial structures were optimized using the PBE
functional\cite{Perdew1996}, augmented with the D3 version of Grimme's
dispersion correction with Becke-Johnson damping\cite{Grimme2010,Grimme2011},
as implemented in the CP2K package\cite{Kuhne2020}. The Goedecker, Teter,
and Hutter (GTH) pseudopotentials\cite{Goedecker1996} and the associated
optimized Gaussian basis sets for molecular
calculations\cite{VandeVondele2007} were used. Specifically, the triple
zeta basis sets (TZVP-MOLOPT-GTH) were used for describing valence
electrons in oxygen and hydrogen, while the double zeta basis sets
(DZVP-MOLOPT-SR-GTH) were used for copper, boron, nitrogen, and carbon,
providing a good balance between accuracy and computational efficiency.
In the Gaussian and Plane Waves (GPW) method, the plane wave cutoff for
the finest grid level and the mapping reference grid were set to 480 and
60~Rydberg, respectively. Periodic boundary conditions were applied in
all three dimensions. For $k$-point sampling, the $\Gamma$-only method
was used during geometrical optimizations, justified by the large
supercell sizes (the lengths of the supercells are all greater than
20~\AA{}), which minimize finite-size effects. The conjugate gradient
algorithm with four steps was employed to accelerate the optimizations.
Specifically, 1) Relaxation of the $z$-coordinates of the 2D ice and the
topmost two Cu(111) slab layers; 2) Full relaxation of all degrees of
the freedom of the 2D ice; 3) Optimization of the $xy$-coordinates of
the capping layer and the topmost two Cu(111) slab layers; 4)
Simultaneous optimization of all degree of the freedoms of the 2D ice
and the topmost two Cu(111) slab layers as well as the $xy$-coordinates
of the capping layer. During these steps, the positions of the bottom
Cu(111) slab layer and the $z$-coordinates of the capping layer were
fixed to mimic the metallic bulk and to simulate experimental conditions
for confinement. Based on the optimized geometries, the stress tensors
were calculated for systems with and without the presence of the 2D ice
monolayer. The true stress tensor of the 2D ice monolayer was obtained
by subtracting the latter from the former. Then, as suggested by the
previous works,\cite{Corsetti2016,Chen2016} the lateral pressure exerted
on the 2D ice was evaluated as
\begin{equation}
P=\frac{L_z}{2h}\left(\sigma_{xx}+\sigma_{yy}\right),
\label{eq_p}
\end{equation}
where $\sigma_{xx}$ and $\sigma_{yy}$ are the diagonal elements of the
stress tensor of the ice layer, $L_z$ is the length of the supercell in
the $z$-direction and $h$ is the confinement width (i.e., the distance
between the h-BN capping layer and the Cu(111) substrate).

\noindent\textbf{Electronic structure calculation}\\
On top of the optimized geometrical structures, the DOS and partial
charge analyses were conducted using the Vienna Ab-initio Simulation
Package (VASP)\cite{Kresse1996}. Calculations employed either the
PBE\cite{Perdew1996} or xPBE\cite{xu2004jcp} functionals via
Libxc\cite{Lehtola2018}. Projector augmented-wave (PAW)
pseudopotentials\cite{Kresse1999} were used to describe core electrons,
while plane-wave basis sets with a kinetic energy cutoff of 400~eV were
used to describe valence electrons. For Brillouin zone integration, a
dense $k$-point mesh was adopted, ensuring a maximum spacing of less than
0.038~\AA{}$^{-1}$, in combination with the tetrahedron method and
Bl\"ochl corrections to achieve high accuracy of electronic structures.
A dipole correction was applied along the $z$-direction to mitigate
artificial electrostatic interactions between periodically repeated
images.

\noindent\textbf{STM simulation}\\
Based on the obtained electronic structures, all STM images were
simulated using the Tersoff-Hamann 
theory\cite{Tersoff1985,duan2020jacs,duan2023jacsau,duan2023jpcl},
assuming for conventional metallic tips under a sample bias of --0.5 V
in constant-current mode. To account for finite mapping resolution, all
calculated STM images were filtered by a Gaussian smearing with a
bandwidth of 0.75~\AA{}. The isovalue settings for different systems
were summarized in Supplementary Table~3. The STM images obtained using
the PBE and xPBE functionals exhibit no observable differences,
demonstrating the robustness of the images.

\section*{\large{Acknowledgements}}

This work was supported by the National Key R\&D Program of China
(2024YFA1208104), the National Natural Science Foundation of China (Nos.
22393911 and 22473028), the Innovation Program for Quantum Science and
Technology (Nos. 2021ZD0303301 and 2021ZD0303305), and the robotic
AI-Scientist platform of Chinese Academy of Science.

\section*{\large{Author contributions}}

S.D. and X.X. conceived and supervised the project. 
B.W. and J.W. performed the calculations.
S.D., X.X. and B.W. analyzed the data and wrote the manuscript.
All authors reviewed, edited, and approved the final manuscript.

\section*{\large{Competing interests}}
The authors declare no competing interests.

\section*{\large{Additional information}}
\begin{addendum}
\item[Supplementary Information] Supplementary information is available in
the online version of the paper. Reprints and permissions information is
available online at www.nature.com/reprints. 
\item[Correspondence] Correspondence and requests for materials should
be addressed to Sai Duan and Xin Xu.
\item[Corresponding Authors] Sai Duan (email: duansai@fudan.edu.cn) and
Xin Xu (email: xxchem@fudan.edu.cn)
\end{addendum}

\section*{\large{References}}
\bibliographystyle{naturemag}
\bibliography{tdw}

\begin{thebibliography}{10}
\expandafter\ifx\csname url\endcsname\relax
  \def\url#1{\texttt{#1}}\fi
\expandafter\ifx\csname urlprefix\endcsname\relax\def\urlprefix{URL }\fi
\providecommand{\bibinfo}[2]{#2}
\providecommand{\eprint}[2][]{\url{#2}}

\bibitem{SzentGyrgyi1971}
\bibinfo{author}{Szent-Gy\"{o}rgyi, A.}
\newblock \bibinfo{title}{Biology and pathology of water}.
\newblock \emph{\bibinfo{journal}{Perspect. Biol. Med.}}
  \textbf{\bibinfo{volume}{14}}, \bibinfo{pages}{239–249}
  (\bibinfo{year}{1971}).

\bibitem{ball2017pnas}
\bibinfo{author}{Ball, P.}
\newblock \bibinfo{title}{Water is an active matrix of life for cell and
  molecular biology}.
\newblock \emph{\bibinfo{journal}{Proc. Natl. Acad. Sci. U.S.A.}}
  \textbf{\bibinfo{volume}{114}}, \bibinfo{pages}{13327--13335}
  (\bibinfo{year}{2017}).

\bibitem{ball2008nature}
\bibinfo{author}{Ball, P.}
\newblock \bibinfo{title}{Water - an enduring mystery}.
\newblock \emph{\bibinfo{journal}{Nature}} \textbf{\bibinfo{volume}{452}},
  \bibinfo{pages}{291–292} (\bibinfo{year}{2008}).

\bibitem{pettersson2016cr}
\bibinfo{author}{Pettersson, L. G.~M.}, \bibinfo{author}{Henchman, R.~H.} \&
  \bibinfo{author}{Nilsson, A.}
\newblock \bibinfo{title}{Water-the most anomalous liquid}.
\newblock \emph{\bibinfo{journal}{Chem. Rev.}} \textbf{\bibinfo{volume}{116}},
  \bibinfo{pages}{7459–7462} (\bibinfo{year}{2016}).

\bibitem{eisenberg2005}
\bibinfo{author}{Eisenberg, D.} \& \bibinfo{author}{Kauzmann, W.}
\newblock \emph{\bibinfo{title}{The Structure and Properties of Water}}
  (\bibinfo{publisher}{Oxford University Press}, \bibinfo{year}{2005}).

\bibitem{brini2017cr}
\bibinfo{author}{Brini, E.} \emph{et~al.}
\newblock \bibinfo{title}{How water’s properties are encoded in its molecular
  structure and energies}.
\newblock \emph{\bibinfo{journal}{Chem. Rev.}} \textbf{\bibinfo{volume}{117}},
  \bibinfo{pages}{12385–12414} (\bibinfo{year}{2017}).

\bibitem{tammann1900ap}
\bibinfo{author}{Tammann, G.}
\newblock \bibinfo{title}{Ueber die grenzen des festen zustandes {IV}}.
\newblock \emph{\bibinfo{journal}{Ann. Phys.}} \textbf{\bibinfo{volume}{307}},
  \bibinfo{pages}{1–31} (\bibinfo{year}{1900}).

\bibitem{Hansen2021}
\bibinfo{author}{Hansen, T.~C.}
\newblock \bibinfo{title}{The everlasting hunt for new ice phases}.
\newblock \emph{\bibinfo{journal}{Nat. Commun.}} \textbf{\bibinfo{volume}{12}},
  \bibinfo{pages}{3161} (\bibinfo{year}{2021}).

\bibitem{salzmann2023science}
\bibinfo{author}{Rosu-Finsen, A.} \emph{et~al.}
\newblock \bibinfo{title}{Medium-density amorphous ice}.
\newblock \emph{\bibinfo{journal}{Science}} \textbf{\bibinfo{volume}{379}},
  \bibinfo{pages}{474--478} (\bibinfo{year}{2023}).

\bibitem{Gopinadhan2019}
\bibinfo{author}{Gopinadhan, K.} \emph{et~al.}
\newblock \bibinfo{title}{Complete steric exclusion of ions and proton
  transport through confined monolayer water}.
\newblock \emph{\bibinfo{journal}{Science}} \textbf{\bibinfo{volume}{363}},
  \bibinfo{pages}{145--148} (\bibinfo{year}{2019}).

\bibitem{Fumagalli2018}
\bibinfo{author}{Fumagalli, L.} \emph{et~al.}
\newblock \bibinfo{title}{Anomalously low dielectric constant of confined
  water}.
\newblock \emph{\bibinfo{journal}{Science}} \textbf{\bibinfo{volume}{360}},
  \bibinfo{pages}{1339--1342} (\bibinfo{year}{2018}).

\bibitem{Nair2012}
\bibinfo{author}{Nair, R.~R.}, \bibinfo{author}{Wu, H.~A.},
  \bibinfo{author}{Jayaram, P.~N.}, \bibinfo{author}{Grigorieva, I.~V.} \&
  \bibinfo{author}{Geim, A.~K.}
\newblock \bibinfo{title}{Unimpeded permeation of water through
  helium-leak-tight graphene-based membranes}.
\newblock \emph{\bibinfo{journal}{Science}} \textbf{\bibinfo{volume}{335}},
  \bibinfo{pages}{442--444} (\bibinfo{year}{2012}).

\bibitem{Neek-Amal2016}
\bibinfo{author}{Neek-Amal, M.}, \bibinfo{author}{Peeters, F.~M.},
  \bibinfo{author}{Grigorieva, I.~V.} \& \bibinfo{author}{Geim, A.~K.}
\newblock \bibinfo{title}{Commensurability effects in viscosity of nanoconfined
  water}.
\newblock \emph{\bibinfo{journal}{ACS Nano}} \textbf{\bibinfo{volume}{10}},
  \bibinfo{pages}{3685--3692} (\bibinfo{year}{2016}).

\bibitem{Guo2025}
\bibinfo{author}{Guo, W.-Q.}, \bibinfo{author}{Deng, J.-W.} \&
  \bibinfo{author}{Wang, B.-B.}
\newblock \bibinfo{title}{Molecular dynamics simulations on heat transport of
  nanoconfined water under electric fields: Effect of nanochannel size}.
\newblock \emph{\bibinfo{journal}{J. Phys. Chem. B}}
  \textbf{\bibinfo{volume}{129}}, \bibinfo{pages}{348--359}
  (\bibinfo{year}{2025}).

\bibitem{Israelachvili1996}
\bibinfo{author}{Israelachvili, J.} \& \bibinfo{author}{Wennerstr{\"o}m, H.}
\newblock \bibinfo{title}{Role of hydration and water structure in biological
  and colloidal interactions}.
\newblock \emph{\bibinfo{journal}{Nature}} \textbf{\bibinfo{volume}{379}},
  \bibinfo{pages}{219--225} (\bibinfo{year}{1996}).

\bibitem{Kavokine2022}
\bibinfo{author}{Kavokine, N.}, \bibinfo{author}{Bocquet, M.-L.} \&
  \bibinfo{author}{Bocquet, L.}
\newblock \bibinfo{title}{Fluctuation-induced quantum friction in nanoscale
  water flows}.
\newblock \emph{\bibinfo{journal}{Nature}} \textbf{\bibinfo{volume}{602}},
  \bibinfo{pages}{84--90} (\bibinfo{year}{2022}).

\bibitem{Nie2010}
\bibinfo{author}{Nie, S.}, \bibinfo{author}{Feibelman, P.~J.},
  \bibinfo{author}{Bartelt, N.~C.} \& \bibinfo{author}{Th\"urmer, K.}
\newblock \bibinfo{title}{Pentagons and heptagons in the first water layer on
  {Pt(111)}}.
\newblock \emph{\bibinfo{journal}{Phys. Rev. Lett.}}
  \textbf{\bibinfo{volume}{105}}, \bibinfo{pages}{026102}
  (\bibinfo{year}{2010}).

\bibitem{Ma2020}
\bibinfo{author}{Ma, R.} \emph{et~al.}
\newblock \bibinfo{title}{Atomic imaging of the edge structure and growth of a
  two-dimensional hexagonal ice}.
\newblock \emph{\bibinfo{journal}{Nature}} \textbf{\bibinfo{volume}{577}},
  \bibinfo{pages}{60--63} (\bibinfo{year}{2020}).

\bibitem{Yang2009}
\bibinfo{author}{Yang, D.-S.} \& \bibinfo{author}{Zewail, A.~H.}
\newblock \bibinfo{title}{Ordered water structure at hydrophobic graphite
  interfaces observed by {4D}, ultrafast electron crystallography}.
\newblock \emph{\bibinfo{journal}{Proc. Natl. Acad. Sci. U.S.A.}}
  \textbf{\bibinfo{volume}{106}}, \bibinfo{pages}{4122--4126}
  (\bibinfo{year}{2009}).

\bibitem{Kimmel2009}
\bibinfo{author}{Kimmel, G.~A.} \emph{et~al.}
\newblock \bibinfo{title}{No confinement needed: Observation of a metastable
  hydrophobic wetting two-layer ice on graphene}.
\newblock \emph{\bibinfo{journal}{J. Am. Chem. Soc.}}
  \textbf{\bibinfo{volume}{131}}, \bibinfo{pages}{12838--12844}
  (\bibinfo{year}{2009}).

\bibitem{Xu2010}
\bibinfo{author}{Xu, K.}, \bibinfo{author}{Cao, P.} \& \bibinfo{author}{Heath,
  J.~R.}
\newblock \bibinfo{title}{Graphene visualizes the first water adlayers on mica
  at ambient conditions}.
\newblock \emph{\bibinfo{journal}{Science}} \textbf{\bibinfo{volume}{329}},
  \bibinfo{pages}{1188--1191} (\bibinfo{year}{2010}).

\bibitem{He2012}
\bibinfo{author}{He, K.~T.}, \bibinfo{author}{Wood, J.~D.},
  \bibinfo{author}{Doidge, G.~P.}, \bibinfo{author}{Pop, E.} \&
  \bibinfo{author}{Lyding, J.~W.}
\newblock \bibinfo{title}{Scanning tunneling microscopy study and
  nanomanipulation of graphene-coated water on mica}.
\newblock \emph{\bibinfo{journal}{Nano Lett.}} \textbf{\bibinfo{volume}{12}},
  \bibinfo{pages}{2665--2672} (\bibinfo{year}{2012}).

\bibitem{Li2015}
\bibinfo{author}{Li, Q.}, \bibinfo{author}{Song, J.},
  \bibinfo{author}{Besenbacher, F.} \& \bibinfo{author}{Dong, M.}
\newblock \bibinfo{title}{Two-dimensional material confined water}.
\newblock \emph{\bibinfo{journal}{Acc. Chem. Res.}}
  \textbf{\bibinfo{volume}{48}}, \bibinfo{pages}{119--127}
  (\bibinfo{year}{2015}).

\bibitem{AlgaraSiller2015}
\bibinfo{author}{Algara-Siller, G.} \emph{et~al.}
\newblock \bibinfo{title}{Square ice in graphene nanocapillaries}.
\newblock \emph{\bibinfo{journal}{Nature}} \textbf{\bibinfo{volume}{519}},
  \bibinfo{pages}{443--445} (\bibinfo{year}{2015}).

\bibitem{Chen2016}
\bibinfo{author}{Chen, J.}, \bibinfo{author}{Schusteritsch, G.},
  \bibinfo{author}{Pickard, C.~J.}, \bibinfo{author}{Salzmann, C.~G.} \&
  \bibinfo{author}{Michaelides, A.}
\newblock \bibinfo{title}{Two dimensional ice from first principles: Structures
  and phase transitions}.
\newblock \emph{\bibinfo{journal}{Phys. Rev. Lett.}}
  \textbf{\bibinfo{volume}{116}}, \bibinfo{pages}{025501}
  (\bibinfo{year}{2016}).

\bibitem{Kapil2022}
\bibinfo{author}{Kapil, V.} \emph{et~al.}
\newblock \bibinfo{title}{The first-principles phase diagram of monolayer
  nanoconfined water}.
\newblock \emph{\bibinfo{journal}{Nature}} \textbf{\bibinfo{volume}{609}},
  \bibinfo{pages}{512--516} (\bibinfo{year}{2022}).

\bibitem{Lin2023}
\bibinfo{author}{Lin, B.}, \bibinfo{author}{Jiang, J.}, \bibinfo{author}{Zeng,
  X.~C.} \& \bibinfo{author}{Li, L.}
\newblock \bibinfo{title}{Temperature-pressure phase diagram of confined
  monolayer water/ice at first-principles accuracy with a machine-learning
  force field}.
\newblock \emph{\bibinfo{journal}{Nat. Commun.}} \textbf{\bibinfo{volume}{14}},
  \bibinfo{pages}{4110} (\bibinfo{year}{2023}).

\bibitem{jiang2024nphys}
\bibinfo{author}{Jiang, J.} \emph{et~al.}
\newblock \bibinfo{title}{Rich proton dynamics and phase behaviours of
  nanoconfined ices}.
\newblock \emph{\bibinfo{journal}{Nat. Phys.}} \textbf{\bibinfo{volume}{20}},
  \bibinfo{pages}{456–464} (\bibinfo{year}{2024}).

\bibitem{Ravindra2024}
\bibinfo{author}{Ravindra, P.}, \bibinfo{author}{Advincula, X.~R.},
  \bibinfo{author}{Schran, C.}, \bibinfo{author}{Michaelides, A.} \&
  \bibinfo{author}{Kapil, V.}
\newblock \bibinfo{title}{Quasi-one-dimensional hydrogen bonding in
  nanoconfined ice}.
\newblock \emph{\bibinfo{journal}{Nat. Commun.}} \textbf{\bibinfo{volume}{15}},
  \bibinfo{pages}{7301} (\bibinfo{year}{2024}).

\bibitem{Zhou2015}
\bibinfo{author}{Zhou, W.} \emph{et~al.}
\newblock \bibinfo{title}{The observation of square ice in graphene
  questioned}.
\newblock \emph{\bibinfo{journal}{Nature}} \textbf{\bibinfo{volume}{528}},
  \bibinfo{pages}{E1--E2} (\bibinfo{year}{2015}).

\bibitem{duan2020jacs}
\bibinfo{author}{Duan, S.}, \bibinfo{author}{Zhang, I.~Y.},
  \bibinfo{author}{Xie, Z.} \& \bibinfo{author}{Xu, X.}
\newblock \bibinfo{title}{Identification of water hexamer on {Cu(111)}
  surfaces}.
\newblock \emph{\bibinfo{journal}{J. Am. Chem. Soc.}}
  \textbf{\bibinfo{volume}{142}}, \bibinfo{pages}{6902--6906}
  (\bibinfo{year}{2020}).

\bibitem{Wickramaratne2018}
\bibinfo{author}{Wickramaratne, D.}, \bibinfo{author}{Weston, L.} \&
  \bibinfo{author}{Van~de Walle, C.~G.}
\newblock \bibinfo{title}{Monolayer to bulk properties of hexagonal boron
  nitride}.
\newblock \emph{\bibinfo{journal}{J. Phys. Chem. C}}
  \textbf{\bibinfo{volume}{122}}, \bibinfo{pages}{25524--25529}
  (\bibinfo{year}{2018}).

\bibitem{Wu2019}
\bibinfo{author}{Wu, H.} \emph{et~al.}
\newblock \bibinfo{title}{Dynamic nanoscale imaging of enriched {CO} adlayer on
  {Pt(111)} confined under {h-BN} monolayer in ambient pressure atmospheres}.
\newblock \emph{\bibinfo{journal}{Nano Res.}} \textbf{\bibinfo{volume}{12}},
  \bibinfo{pages}{85--90} (\bibinfo{year}{2019}).

\bibitem{Kirchhoff2022}
\bibinfo{author}{Kirchhoff, A.}, \bibinfo{author}{Deilmann, T.},
  \bibinfo{author}{Kr\"uger, P.} \& \bibinfo{author}{Rohlfing, M.}
\newblock \bibinfo{title}{Electronic and optical properties of a hexagonal
  boron nitride monolayer in its pristine form and with point defects from
  first principles}.
\newblock \emph{\bibinfo{journal}{Phys. Rev. B}}
  \textbf{\bibinfo{volume}{106}}, \bibinfo{pages}{045118}
  (\bibinfo{year}{2022}).

\bibitem{Tersoff1985}
\bibinfo{author}{Tersoff, J.} \& \bibinfo{author}{Hamann, D.~R.}
\newblock \bibinfo{title}{Theory of the scanning tunneling microscope}.
\newblock \emph{\bibinfo{journal}{Phys. Rev. B}} \textbf{\bibinfo{volume}{31}},
  \bibinfo{pages}{805--813} (\bibinfo{year}{1985}).

\bibitem{duan2023jacsau}
\bibinfo{author}{Duan, S.}, \bibinfo{author}{Tian, G.} \& \bibinfo{author}{Xu,
  X.}
\newblock \bibinfo{title}{A general framework of scanning tunneling microscopy
  based on {Bardeen's} approximation for isolated molecules}.
\newblock \emph{\bibinfo{journal}{JACS Au}} \textbf{\bibinfo{volume}{3}},
  \bibinfo{pages}{86--92} (\bibinfo{year}{2023}).

\bibitem{duan2023jpcl}
\bibinfo{author}{Duan, S.} \& \bibinfo{author}{Xu, X.}
\newblock \bibinfo{title}{Accurate simulations of scanning tunneling
  microscope: Both tip and substrate states matter}.
\newblock \emph{\bibinfo{journal}{J. Phys. Chem. Lett.}}
  \textbf{\bibinfo{volume}{14}}, \bibinfo{pages}{6726--6735}
  (\bibinfo{year}{2023}).

\bibitem{zhu2025jacsau}
\bibinfo{author}{Zhu, Y.} \emph{et~al.}
\newblock \bibinfo{title}{Reconstructing pristine molecular orbitals from
  scanning tunneling microscope images via artificial intelligence approaches}.
\newblock \emph{\bibinfo{journal}{JACS Au}} \textbf{\bibinfo{volume}{5}},
  \bibinfo{pages}{3163--3170} (\bibinfo{year}{2025}).

\bibitem{Geim2007}
\bibinfo{author}{Geim, A.~K.} \& \bibinfo{author}{Novoselov, K.~S.}
\newblock \bibinfo{title}{The rise of graphene}.
\newblock \emph{\bibinfo{journal}{Nat. Mater.}} \textbf{\bibinfo{volume}{6}},
  \bibinfo{pages}{183--191} (\bibinfo{year}{2007}).

\bibitem{Abergel2010}
\bibinfo{author}{Abergel, D.}, \bibinfo{author}{Apalkov, V.},
  \bibinfo{author}{Berashevich, J.}, \bibinfo{author}{Ziegler, K.} \&
  \bibinfo{author}{Chakraborty, T.}
\newblock \bibinfo{title}{Properties of graphene: a theoretical perspective}.
\newblock \emph{\bibinfo{journal}{Adv. Phys.}} \textbf{\bibinfo{volume}{59}},
  \bibinfo{pages}{261--482} (\bibinfo{year}{2010}).

\bibitem{gross2009science}
\bibinfo{author}{Gross, L.}, \bibinfo{author}{Mohn, F.}, \bibinfo{author}{Moll,
  N.}, \bibinfo{author}{Liljeroth, P.} \& \bibinfo{author}{Meyer, G.}
\newblock \bibinfo{title}{The chemical structure of a molecule resolved by
  atomic force microscopy}.
\newblock \emph{\bibinfo{journal}{Science}} \textbf{\bibinfo{volume}{325}},
  \bibinfo{pages}{1110--1114} (\bibinfo{year}{2009}).

\bibitem{duan2015jacs}
\bibinfo{author}{Duan, S.} \emph{et~al.}
\newblock \bibinfo{title}{Theoretical modeling of plasmon-enhanced {Raman}
  images of a single molecule with subnanometer resolution}.
\newblock \emph{\bibinfo{journal}{J. Am. Chem. Soc.}}
  \textbf{\bibinfo{volume}{137}}, \bibinfo{pages}{9515--9518}
  (\bibinfo{year}{2015}).

\bibitem{duan2016angew}
\bibinfo{author}{Duan, S.}, \bibinfo{author}{Tian, G.} \& \bibinfo{author}{Luo,
  Y.}
\newblock \bibinfo{title}{Visualization of vibrational modes in real space by
  tip‐enhanced non‐resonant {Raman} spectroscopy}.
\newblock \emph{\bibinfo{journal}{Angew. Chem. Int. Ed.}}
  \textbf{\bibinfo{volume}{55}}, \bibinfo{pages}{1041–1045}
  (\bibinfo{year}{2016}).

\bibitem{duan2019jacs}
\bibinfo{author}{Duan, S.}, \bibinfo{author}{Rinkevicius, Z.},
  \bibinfo{author}{Tian, G.} \& \bibinfo{author}{Luo, Y.}
\newblock \bibinfo{title}{Optomagnetic effect induced by magnetized nanocavity
  plasmon}.
\newblock \emph{\bibinfo{journal}{J. Am. Chem. Soc.}}
  \textbf{\bibinfo{volume}{141}}, \bibinfo{pages}{13795--13798}
  (\bibinfo{year}{2019}).

\bibitem{qiu2022jacsau}
\bibinfo{author}{Qiu, F.} \emph{et~al.}
\newblock \bibinfo{title}{Optical images of molecular vibronic couplings from
  tip-enhanced fluorescence excitation spectroscopy}.
\newblock \emph{\bibinfo{journal}{JACS Au}} \textbf{\bibinfo{volume}{2}},
  \bibinfo{pages}{150--158} (\bibinfo{year}{2022}).

\bibitem{pagliero2025nl}
\bibinfo{author}{Pagliero, D.} \emph{et~al.}
\newblock \bibinfo{title}{Slow water in engineered nanochannels revealed by
  color-center-enabled sensing}.
\newblock \emph{\bibinfo{journal}{Nano Lett.}} \textbf{\bibinfo{volume}{25}},
  \bibinfo{pages}{9960--9966} (\bibinfo{year}{2025}).

\bibitem{zheng2024arxiv}
\bibinfo{author}{Zheng, W.} \emph{et~al.}
\newblock \bibinfo{title}{Observation of liquid-solid transition of
  nanoconfined water at ambient temperature} (\bibinfo{year}{2024}).
\newblock \bibinfo{note}{ArXiv:2412.15001}.

\bibitem{Perdew1996}
\bibinfo{author}{Perdew, J.~P.}, \bibinfo{author}{Burke, K.} \&
  \bibinfo{author}{Ernzerhof, M.}
\newblock \bibinfo{title}{Generalized gradient approximation made simple}.
\newblock \emph{\bibinfo{journal}{Phys. Rev. Lett.}}
  \textbf{\bibinfo{volume}{77}}, \bibinfo{pages}{3865--3868}
  (\bibinfo{year}{1996}).

\bibitem{Grimme2010}
\bibinfo{author}{Grimme, S.}, \bibinfo{author}{Antony, J.},
  \bibinfo{author}{Ehrlich, S.} \& \bibinfo{author}{Krieg, H.}
\newblock \bibinfo{title}{A consistent and accurate ab initio parametrization
  of density functional dispersion correction ({DFT-D}) for the 94 elements
  {H-Pu}}.
\newblock \emph{\bibinfo{journal}{J. Chem. Phys.}}
  \textbf{\bibinfo{volume}{132}}, \bibinfo{pages}{154104}
  (\bibinfo{year}{2010}).

\bibitem{Grimme2011}
\bibinfo{author}{Grimme, S.}, \bibinfo{author}{Ehrlich, S.} \&
  \bibinfo{author}{Goerigk, L.}
\newblock \bibinfo{title}{Effect of the damping function in dispersion
  corrected density functional theory}.
\newblock \emph{\bibinfo{journal}{J. Comput. Chem.}}
  \textbf{\bibinfo{volume}{32}}, \bibinfo{pages}{1456--1465}
  (\bibinfo{year}{2011}).

\bibitem{Kuhne2020}
\bibinfo{author}{K\"uhne, T.~D.} \emph{et~al.}
\newblock \bibinfo{title}{{CP2K}: An electronic structure and molecular
  dynamics software package - quickstep: Efficient and accurate electronic
  structure calculations}.
\newblock \emph{\bibinfo{journal}{J. Chem. Pnys.}}
  \textbf{\bibinfo{volume}{152}}, \bibinfo{pages}{194103}
  (\bibinfo{year}{2020}).

\bibitem{Goedecker1996}
\bibinfo{author}{Goedecker, S.}, \bibinfo{author}{Teter, M.} \&
  \bibinfo{author}{Hutter, J.}
\newblock \bibinfo{title}{Separable dual-space gaussian pseudopotentials}.
\newblock \emph{\bibinfo{journal}{Phys. Rev. B}} \textbf{\bibinfo{volume}{54}},
  \bibinfo{pages}{1703--1710} (\bibinfo{year}{1996}).

\bibitem{VandeVondele2007}
\bibinfo{author}{VandeVondele, J.} \& \bibinfo{author}{Hutter, J.}
\newblock \bibinfo{title}{Gaussian basis sets for accurate calculations on
  molecular systems in gas and condensed phases}.
\newblock \emph{\bibinfo{journal}{J. Chem. Phys.}}
  \textbf{\bibinfo{volume}{127}}, \bibinfo{pages}{114105}
  (\bibinfo{year}{2007}).

\bibitem{Corsetti2016}
\bibinfo{author}{Corsetti, F.}, \bibinfo{author}{Matthews, P.} \&
  \bibinfo{author}{Artacho, E.}
\newblock \bibinfo{title}{Structural and configurational properties of
  nanoconfined monolayer ice from first principles}.
\newblock \emph{\bibinfo{journal}{Sci. Rep.}} \textbf{\bibinfo{volume}{6}},
  \bibinfo{pages}{18651} (\bibinfo{year}{2016}).

\bibitem{Kresse1996}
\bibinfo{author}{Kresse, G.} \& \bibinfo{author}{Furthm\"uller, J.}
\newblock \bibinfo{title}{Efficient iterative schemes for ab initio
  total-energy calculations using a plane-wave basis set}.
\newblock \emph{\bibinfo{journal}{Phys. Rev. B}} \textbf{\bibinfo{volume}{54}},
  \bibinfo{pages}{11169--11186} (\bibinfo{year}{1996}).

\bibitem{xu2004jcp}
\bibinfo{author}{Xu, X.} \& \bibinfo{author}{Goddard~III, W.~A.}
\newblock \bibinfo{title}{The extended {Perdew-Burke-Ernzerhof} functional with
  improved accuracy for thermodynamic and electronic properties of molecular
  systems}.
\newblock \emph{\bibinfo{journal}{J. Chem. Phys.}}
  \textbf{\bibinfo{volume}{121}}, \bibinfo{pages}{4068--4082}
  (\bibinfo{year}{2004}).

\bibitem{Lehtola2018}
\bibinfo{author}{Lehtola, S.}, \bibinfo{author}{Steigemann, C.},
  \bibinfo{author}{Oliveira, M. J.~T.} \& \bibinfo{author}{Marques, M. A.~L.}
\newblock \bibinfo{title}{Recent developments in libxc — a comprehensive
  library of functionals for density functional theory}.
\newblock \emph{\bibinfo{journal}{SoftwareX}} \textbf{\bibinfo{volume}{7}},
  \bibinfo{pages}{1--5} (\bibinfo{year}{2018}).

\bibitem{Kresse1999}
\bibinfo{author}{Kresse, G.} \& \bibinfo{author}{Joubert, D.}
\newblock \bibinfo{title}{From ultrasoft pseudopotentials to the projector
  augmented-wave method}.
\newblock \emph{\bibinfo{journal}{Phys. Rev. B}} \textbf{\bibinfo{volume}{59}},
  \bibinfo{pages}{1758--1775} (\bibinfo{year}{1999}).

\end{thebibliography}

\renewcommand{\thefigure}{\arabic{figure}}
\setcounter{figure}{0}
\renewcommand{\figurename}{Extended Data Figure}

\clearpage
\newpage

\begin{figure}
\centering
\includegraphics{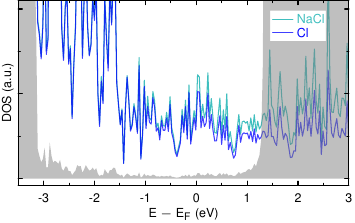}
\caption{\textbf{Detailed Projected DOS for the NaCl monolayer capped by
h-BN}. The cyan and blue lines represent the projected DOS contributed
from the total NaCl layer and that from the Cl atoms, respectively. The
projected DOS of the h-BN (gray region) capping layer is also depicted
for comparison.}
\label{fig_ex1}
\end{figure}

\clearpage
\newpage

\begin{figure}
\centering
\includegraphics{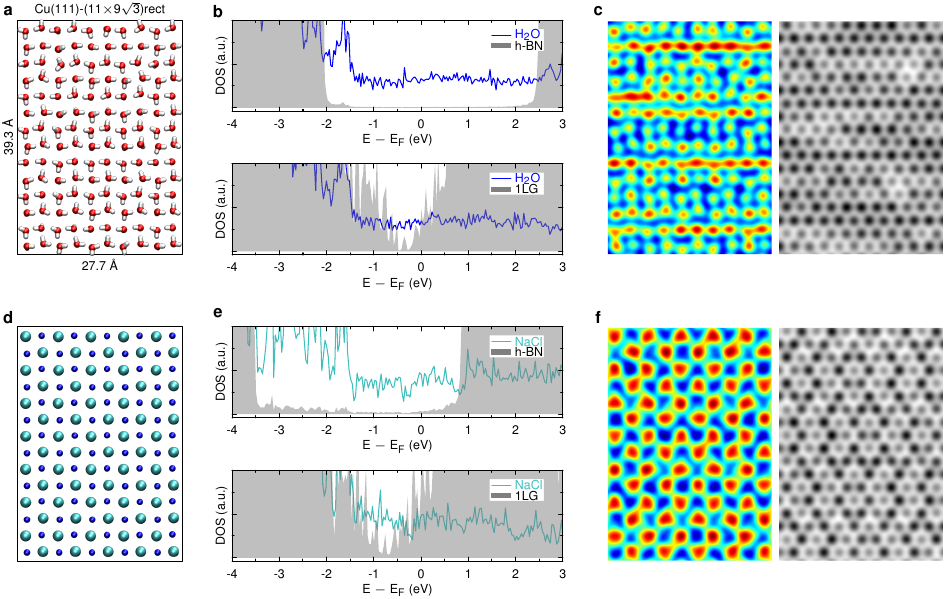}
\caption{\textbf{Identification of an alternative square 2D ice}.
\textbf{a}, Top view of the optimized geometry of the square 2D ice that
has the primitive water unit cell reported in Ref.~\citenum{Chen2016},
confined between a single h-BN layer and Cu(111) surface with a
confinement width of 5.5~\AA{}. The lattice constants (in~\AA{}) of the
heterogeneous layers and the supercell of Cu(111) are labeled.
\textbf{b}, Projected density of state (DOS) for the alternative square
2D ice systems with a single h-BN layer (top) and a single graphene
layer (1LG, bottom) as the capping layer. \textbf{c}, Simulated STM
images for confined square 2D ice with h-BN (left) and graphene (right)
as the capping layer. \textbf{d}, Top view after replacing water with
NaCl in panel (\textbf{a}). \textbf{e}, Projected DOS for the NaCl
systems capped by h-BN (top) and graphene layer (1LG, bottom) capping
NaCl. \textbf{f}, Simulated STM images of confined NaCl with a single
h-BN layer (left) and a single graphene layer (right) as the capping
layer. All STM simulations were performed at a sample bias of --0.5~V.}
\label{fig_ex2}
\end{figure}

\clearpage
\newpage

\begin{figure}
\centering
\includegraphics{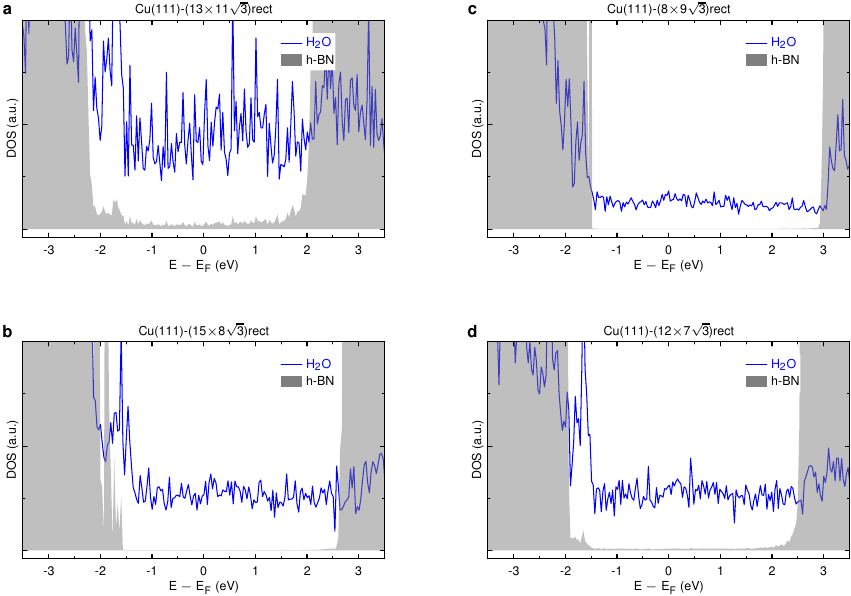}
\caption{\textbf{Projected DOS for other 2D ice structures}. Calculated
projected DOS of the zigzag (\textbf{a}), square/zigzag coexisting
(\textbf{b}), pentagonal with rectangle supercell (\textbf{c}), and
pentagonal with square supercell (\textbf{d}) 2D ice confined between
the h-BN and Cu(111) surface. The blue line and gray region represent
the projected DOS contributed from the 2D ice layer and that from the
h-BN capping layer, respectively. The supercells of Cu(111) are also
indicated.}
\label{fig_ex3}
\end{figure}

\end{document}